\documentstyle[pra,aps,epsf,preprint]{revtex} 
						  
\begin{document}

\bibliographystyle{prsty} 

\title{Photon Production in Heavy-Ion Collisions close to the Pion Threshold}

\author{G.~Mart\'\i nez$^1$\thanks{Present address: SUBATECH, 44307 Nantes, France},  
L.~Aphecetche$^{1*}$, 
Y.~Charbonnier$^1$\thanks{Present address: ATOS, 92142 Clamart, France},   
H.~Delagrange$^{1*}$, 
D.~d'Enterria$^{1*}$,
T.~Matulewicz$^{1,7}$, 
Y.~Schutz$^{1*}$, 
R.~Turrisi$^1$\thanks{Present address: Istituto di F\'{\i}sica dell'Universita, 95129 Catania, Italy} , 
M.~Appenheimer$^2$,  V.~Metag$^2$, R.~Novotny$^2$, 
H.~Str\"oher$^2$\thanks{Present address: Institut f\"ur Kernphysik, 52425 J\"ulich, Germany},
A.R.~Wolf$^2$, M.~Wolf$^2$, J.~Wei\ss$^2$, R.~Averbeck$^3$, 
S.~Hlav\'a\v c$^3$\thanks{Permanent address: Institute 
of Physics, Slovak Academy of Sciences, Bratislava, Slovak Republic},  
R.~Holzmann$^3$, F.~Lef\`evre$^{3*}$,
R.S.~Simon$^3$, R.~Stratmann$^3$\thanks{Present address: CompuMED GmbH, 
56070 Koblenz, Germany}, F.~Wissmann$^3$, M.~Hoefman$^4$,
M.J.~van Goethem$^4$, H.~L\"ohner$^4$, R.W~Ostendorf$^4$,
R.H.~Siemssen$^4$\thanks{Present address: Argonne National Laboratory,
Argonne, IL60439, USA}, P.~Vogt$^4$, H.W.~Wilschut$^4$,
J.~D\'\i az$^5$, A.~Mar\'\i n$^5$\thanks{Present address: GSI, 64291 
Darmstadt, Germany}, A.~Kugler$^6$, P.~Tlusty$^6$, V.~Wagner$^6$ and A.~D\"oppenschmidt$^8$
}

\address{$^1$Grand Acc\'el\'erateur National d'Ions Lourds, 
		    14076 Caen, France}
\address{$^2$II. Physikalisches Institut, 
                 Universit\"at Gie\ss en, 35392 Gie\ss en, Germany}	
\address{$^3$Gesellschaft f\"ur Schwerionenforschung,
                  64291 Darmstadt, Germany}			 		
\address{$^4$Kernfysisch Versneller Instituut,
                   9747 AA Groningen, The Netherlands}
\address{$^5$Instituto de F\'{\i}sica Corpuscular, 
                  46100-Burjassot, Spain} 		   
\address{$^6$Institute of Nuclear Physics, 250 68 \v Re\v z, Czech Republic}
\address{$^7$Institute of Experimental Physics, Warsaw University, 00-681, 
              Warsaw, Poland}
\address{$^8$Institut f\"ur Kernphysik, Universit\"at Frankfurt, 60486 Frankfurt-am-Main, Germany}
\date{\today}
\maketitle	  
\begin{abstract}

We report on a measurement of  hard photons ($E_\gamma>30$ MeV) in the
reaction Ar+Ca at 180$A$ MeV at an energy in which photons from the decay
of $\pi^\circ$ mesons are dominating. 
Simultaneous measurement with the TAPS spectrometer of the photon spectrum and
photon-photon coincidences used for the identification of $\pi^\circ$ enabled the
subtraction of $\pi^\circ$ contribution.
The resulting photon spectrum exhibits an exponential shape with an inverse
slope of $E_0=(53\pm2_{(stat)}~^{-5}_{+8}~_{(syst)})$ MeV. The photon
multiplicity equal to $(1.21\pm 0.03_{(stat)}~^{+0.3}_{-0.2}~_{(syst)})\cdot
10^{-2}$ is roughly one order of magnitude larger than  the value extrapolated
from existing systematics. This enhancement of the hard photon production is 
attributed to a strong increase in the contribution of secondary $np$
collisions to the total photon yield. 
We conclude that, on average, the number of $np$ collisions 
which contribute to the hard photon production is 
7 times larger than the number of first chance $np$ collisions
in the reaction Ar+Ca at 180A MeV.

\end{abstract}

Hard photons ($E_\gamma>30$ MeV) produced in heavy-ion collisions 
have been proposed and since extensively exploited \cite{Nife89,Cass90,Schu97} as probes of
excited nuclear matter formed in the early stages of nucleus-nucleus reactions.
In contrast  to hadronic probes photons interact only weakly with the
surrounding nuclear medium and therefore convey unperturbed information of the
photon-creation phases of the reaction. Insight on the dynamical evolution of nuclear
reactions leading to the formation of a hot and dense nuclear system 
could be gained by studing the hard photon production in inclusive as well as exclusive 
measurements \cite{Schu97}.
Such measurements were up to now limited to  bombarding energies below
125$A$ MeV \cite{Clay90} because of the fast increase of the pion
production cross-section, resulting in  a strong background of
photons from the dominant electromagnetic $\pi^\circ$ decay 
($\pi^\circ\rightarrow\gamma\gamma$).  

The apparent saturation of the photon production cross-section with
bombarding energy, as deduced from the commonly adopted extrapolation of 
hard photon systematics \cite{Cass90,Polt95} together with the
overabundant photon yield from the neutral meson decay seemed to place
the hard photon detection at higher bombarding energies beyond experimental
reach. 
Investigation of hard photon production at relativistic energies
would, however, provide a new insight to phenomena such
as the formation phase of resonant matter excited in central collisions
\cite{Meta96} or the thermalization process of highly excited projectile- and
target-like fragments created in semi-peripheral reactions \cite{Poch95}. 
To explore this direction, we have investigated the production of hard photons
in the reaction $^{40}$Ar+$^{40}$Ca at 180$A$ MeV using the TAPS photon
spectrometer. The hard-photon spectrum was obtained after careful subtraction
of the $\pi^\circ$-decay contribution from the inclusive photon spectrum. We
find that the photon multiplicity is noticeably larger than the value
extrapolated from the hard photon systematics \cite{Cass90,Polt95}.  
This enhancement of the hard photon yield is ascribed to the increasing 
importance of secondary $np$ collisions to the production of hard photons 
\cite{Schu94,Mart95}.

The $^{40}$Ar beam was delivered by the heavy-ion synchrotron SIS at GSI,
Darmstadt, with an average intensity of $5\times 10^8$ particles in spills of
about 9 s. The total number of accumulated beam particles was $4\times
10^{13}$. The natural calcium target was 320 mg/cm$^2$ thick. A Start
Detector (SD) \cite{Wolf94} consisting of 32 NE102 plastic-scintillators
surrounding the target at a distance of 101 mm  signaled the
occurrence of a reaction and delivered the start signal for time-of-flight
measurements. The SD efficiency averaged over the impact parameter,
$<\epsilon_s>_b=0.56$, was estimated from GEANT simulations
\cite{KANE96} using the FREESCO event generator \cite{Fai86}. The
average SD efficiency when a hard photon is produced was calculated by
weighting the impact-parameter distribution with the distribution of the number
of participant nucleons $A_{\rm part}(b)$ \cite{Nife85}. We found
$<\epsilon_s>_{A_{part}(b)}=0.90$. 
As one can expect, $<\epsilon_s>_{A_{part}(b)}$ slightly increases when 
the impact-parameter distribution is weighted by $[A_{\rm part}(b)]^\alpha$
($\alpha > 1$), being the induced systematic error lower than 10\%.
The reaction rate measured with the SD was
$p_s=6.9\times 10^{-3}$ per beam particle. The TAPS multidetector \cite{Novo91}
was used for photon detection. The 384 TAPS modules (a BaF$_2$ crystal
associated  with a plastic veto scintillator)  were assembled in six
blocks of $8\times 8$ detectors and mounted in two symmetric towers of three
blocks each \cite{Mart97}. The towers were positioned at $\theta=\pm 70^\circ$,
80 cm away from the target, on each side of the beam direction. The photon
trigger was defined by requiring a neutral hit in the TAPS multidetector, with
a deposited energy of at least 10 MeV  in the BaF$_2$ crystals without a
coincident hit in the corresponding veto detector, validated with a reaction
trigger given by the SD.

Photons were discriminated against hadrons through time of flight,
BaF$_2$ pulse-shape analysis, and by requiring an anti-coincidence with the corresponding
charged-particle veto detector \cite{Mart97}. The energy calibration of the
BaF$_2$ crystals was based on the energy loss of minimum ionizing 
cosmic-ray muons.
Photon momenta were reconstructed from the electromagnetic shower using a
clustering algorithm \cite{Marq95}. The photon (E$_\gamma>$ 30 MeV) efficiency
of the detection system, $\epsilon_\gamma=12.3\%$, was calculated with the
GEANT package assuming that photons are emitted from a moving source at the
velocity of the nucleon-nucleon ($NN$) center-of-mass. 
The angular distribution was taken as the sum of an isotropic and a dipolar term \cite{Nife89}.
The uncertainty on the dipolar contribution induces
a systematic error in the photon efficiency of 5\%.
The measured photon
spectrum at mid-rapidity (Fig.1) exhibits a convex shape due to the contribution of the 
$\pi^\circ$ electromagnetic decay. Contributions from
heavier mesons or baryonic resonances can be safely neglected below the pion
threshold energy. 
The total measured photon multiplicity for energies larger than 30 MeV is
$(0.025\pm 0.003)$. 

To calculate the $\pi^\circ$ contribution to the measured photon spectrum (Fig.1), 
the $m_t$ distribution of $\pi^\circ$ (where $m_t$ is the transverse mass, $m_t=\sqrt{p_t^2+m^2}$) 
was measured at mid-rapidity by invariant mass reconstruction of photon pairs (Fig.2). 
The $m_t$ 
distribution exhibits an exponential shape in the $m_t$ range 180 to 400 MeV/$c^2$
with an inverse slope $T=(24 \pm 1)$ MeV and deviates from the exponential behaviour at $m_t$ values below
180 MeV/c$^2$ due to the energy-dependent absorption \cite{Maye93} and $\pi^\circ$ rescattering \cite{Holz96}.
However, because of the limited $\pi^\circ$ rapidity acceptance of TAPS, one needs to extrapolate the measured
$\pi^\circ$ distribution, $dM_{\pi^\circ}/dm_t$, to the full solid angle. An anisotropic pion
emission in the $NN$ center-of-mass was assumed and the TAPS response was 
calculated from GEANT simulations.
The  anisotropy of the $\pi^\circ$ angular distribution is parameterized as
$(1+b_{ani}\cos^2{\theta_{CM}})$. For light systems, the anisotropy $b_{ani}$
has been found to be  close to 1.0 at intermediate and relativistic
bombarding energies \cite{Schu94b,Naga81}. In the only existing measurement at 200$A$
MeV \cite{Mill87} performed with a heavy system, a flat angular distribution
was found for charged pions. Since the anisotropy cannot  be determined from our
data, we have considered, as suggested by the systematics \cite{Schu94b,Naga81,Mill87},
$b_{ani}=1.0$ for the determination of the $\pi^\circ$ 
contribution to the photon spectrum (Fig.1). Systematic errors
due to the uncertainty on the anisotropy were estimated by setting
$b_{ani}=0.0$ and $b_{ani}=2.0$, respectively, and these are shown
in Fig.1 as well. The systematic error also includes the uncertainties on
SD and  photon efficiencies. 

After subtraction of the $\pi^\circ$ contribution to the photon yield, 
the inclusive hard photon
spectrum (Fig.3) exhibits the usual exponential shape which can be parametrized
as: 
\begin{equation}
\frac{dM_\gamma}{dE_\gamma}=\frac{M_\gamma}{E_0} 
\exp{\left(\frac{E_s-E_\gamma}{E_0}\right)}, 
\end{equation}
where $E_s=30$ MeV is the threshold energy of measured photons. 
The fitted inverse slope in the energy range from $E_\gamma$=30 to 150 MeV is $E_0=(53\pm2_{(stat)}~^{-5}_{+8}~_{(syst)})$ MeV. 
The experimental photon multiplicity per nuclear reaction ($E_\gamma \geq
30$ MeV) amounts to $M_\gamma=(1.21\pm 0.03_{(stat)}~^{+0.3}_{-0.2}~_{(syst)}
)\cdot 10^{-2}$.

The systematics of hard photon production ($E_\gamma>E_s=30~MeV$) \cite{Cass90,Polt95} at intermediate energies
points to individual first chance  $np$
bremsstrahlung as the main mechanism behind their production, although
secondary contributions to the photon production beyond first chance  $np$
bremsstrahlung have also been found \cite{Schu94,Mart95,Gudi96}. 
The hard photon
multiplicity is found to scale as  $M_\gamma = <N_{np}^{1st-chance}>
P_\gamma^{1,np}$, where $<N_{np}^{1st-chance}>$ is the number  of first chance
$np$ collisions averaged over the impact parameter \cite{Cass90} and
$P_\gamma^{1,np}$ is the hard photon probability per first chance $np$
collision. Moreover, the photon probability scales with the inverse slope of
the hard photon spectrum as $P_\gamma^{1,np}=M_0 \exp{(-E_s/E_0)}$, 
leading to a saturation of the hard photon probability at beam
energies well above the hard photon threshold. However, the present measurement
of the hard photon multiplicity 
is almost one order of
magnitude larger than the values extrapolated from the previously discussed
systematics: $(1.6\pm 0.2)\cdot 10^{-3}$ \cite{Cass90} and $(1.35\pm0.05)\cdot
10^{-3}$ \cite{Polt95}. In contrast the measured inverse slope of the photon
spectrum,
$E_0=(53\pm2_{(stat)}~^{-5}_{+8}~_{(syst)})$ MeV,
remains compatible with the systematics:  $E_0=46\pm 4$ MeV \cite{Cass90}
and $E_0=53\pm 7$ MeV \cite{Polt95}. 

To gain more insight into  the origin of the enhancement of the photon
production, we have examined, within a simple model, the evolution of
$P_\gamma^{1,np}$ as a function of the beam energy
(Fig.4). The hard photon probability from individual $np$ reactions is estimated as
\begin{equation}
P_\gamma^{1,np} = \frac{\sigma_{np\rightarrow np\gamma} (\sqrt{s})  }{  \sigma_{np} (\sqrt{s})   }
\end{equation}
$\sigma_{np\rightarrow np\gamma}(\sqrt{s})$ is obtained adopting the parameterization of the elementary
mechanism $p+n \rightarrow p+n+\gamma$ proposed in  reference \cite{Scha91}, 
$\sigma_{np} (\sqrt{s})$ is the total np cross section \cite{Gale87}, and $\sqrt{s}$ is the
available energy in the $np$ center of mass, taking into account the coupling of the 
nucleon Fermi momenta with the beam momentum. 
The local Thomas-Fermi approximation and a Wood-Saxon
parameterization of the nuclear density have been assumed, with a density in
the center of the nucleus equal to the saturation density, $\rho_0=0.17$
fm$^{-3}$. 
The calculated value of $P_\gamma^{1,np}$ (Fig.4) roughly reproduces the systematics of
$P_\gamma^{1,np}$  at energies above the hard photon production threshold ($E_{th}\sim 2E_s=60A$ MeV), and
corroborates the saturation of $P_\gamma^{1,np}$ at higher beam energies.  
Assuming that only neutron-proton bremsstrahlung is at the origin of hard photons,
the measured enhancement
of the hard photon multiplicity, should be searched  for in secondary $np$
collisions having still enough energy to produce additonal photons with 
similar probability. Within this interpretation, the measured photon
multiplicity can then be related to the total number of $np$ collisions in the
reaction, which can be calculated through: 
\begin{equation}
<N_{np}> = \frac{M_\gamma}{P_\gamma^{1,np}}
\end{equation}
where $M_\gamma$ is the measured hard photon multiplicity and $P_\gamma^{1,np}$
is the calculated photon probability per $np$ collisions (Eq.2).
Using this expression, we obtain $<N_{np}>=27\pm 7$ for the reaction Ar+Ca at 180$A$ MeV, 
which is roughly 7 times larger than the number of first
chance $np$ collisions, $<N_{np}^{1st-chance}>=4$. 
This result indicates that about 85\% of the hard photon yield is produced in secondary $np$
collisions. 
Dynamical phase-space calculations \cite{Cass99} for the studied system,
although, slightly overpredicting the measured hard photon multiplicity (Fig.3), confirm 
that about 78\% of the hard photon yield is produced in secondary $np$ collisions.

In summary, we have measured the hard photon spectrum in the reaction
Ar+Ca at 180$A$ MeV, which is for the first time that these have been
measured at beam energies close to the pion threshold. The
hard photon spectrum exhibits an exponential shape with an inverse slope
$E_0=(53\pm2_{(stat)}~^{-5}_{+8}~_{(syst)})$ MeV. The measured hard photon
multiplicity $M_\gamma=(1.21\pm 0.03_{(stat)}~^{+0.3}_{-0.2}~_{(syst)}
)\cdot 10^{-2}$ 
is almost one order of magnitude
larger than values obtained from the systematics \cite{Cass90,Polt95}.
We interpret the observed enhancement of the hard photon
production  as  being  due to a strong increase of the contribution of
secondary $np$ collisions to the photon yield  at energies well above the hard
photon production threshold (60$A$ MeV).
We deduce that $(27\pm 7)$ $np$ collisions, on average, take part in the hard photon 
production for the reaction Ar+Ca at 180A MeV.

We would like to thank the GSI technical staff for their help
and delivery of the high quality Ar beam required for the present
measurement,
and Prof. W.Cassing for providing us with recent theoretical results.
This work was in part supported
by the Institut National de Physique Nucl\'eaire et de Physique
des Particules, France,
by the Commissariat de l'Energie Atomique, France,
by the Dutch Stichting voor Fundamenteel Onderzoek der Materie,
The Netherlands,
by the Direcci\'on General de Investigaci\'on Cient\'{\i}fica y
T\'ecnica, Spain, 
by the Generalitat Valenciana, Spain,
by the Bundesministerium f\"ur Bildung, Wissenschaft, Forschung und
Technologie, Germany,
by the Deutsche Forschungsgemeinschaft, Germany,
by the Grant Agency of the Czech Republic,
and by the the European Union HCM network contract HRXCT94066.

\bibliography{taps}

\newpage

\begin{figure}
  \begin{center}
   \mbox{\epsfxsize=15cm   \epsfbox{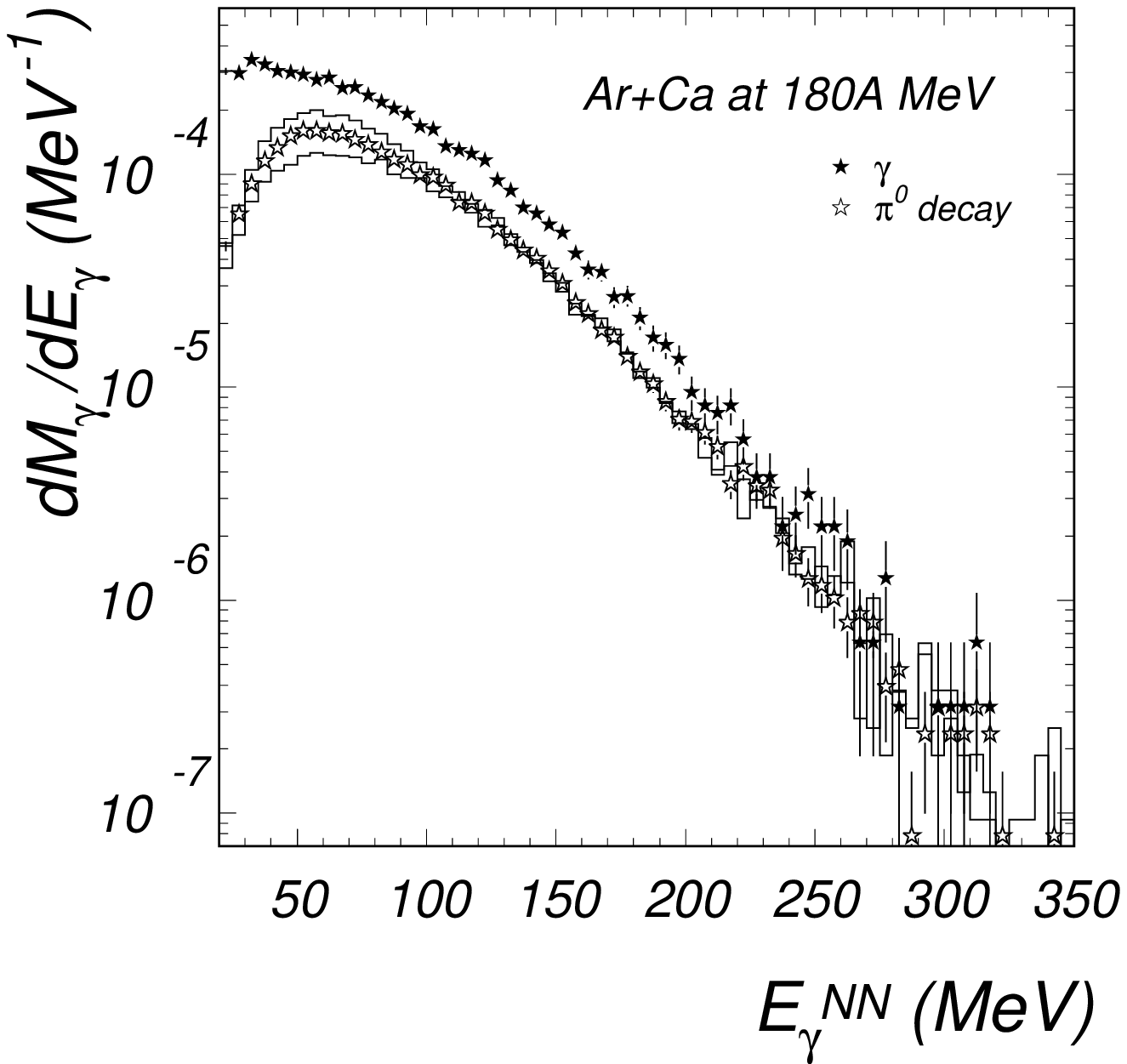}}
  \end{center}
  \label{photon_2}
  \caption{\it Inclusive photon {\it spectrum } 
measured in the $NN$ center-of-mass (full symbols)
  for the reaction Ar+Ca at 180$A$ MeV.
  The contribution from the two-photon decay of $\pi^\circ$ mesons has been deduced by
  extrapolating the measured $\pi^\circ$ yield to the full solid angle assuming the measured transverse mass distribution
  with an anisotropy value $b_{ani}$=1.0 (open symbols). Solid lines delimit the range of systematic 
  errors induced by the uncertainty on this extrapolation.}
\end{figure}

\begin{figure}
  \begin{center}
   \mbox{\epsfxsize=15cm   \epsfbox{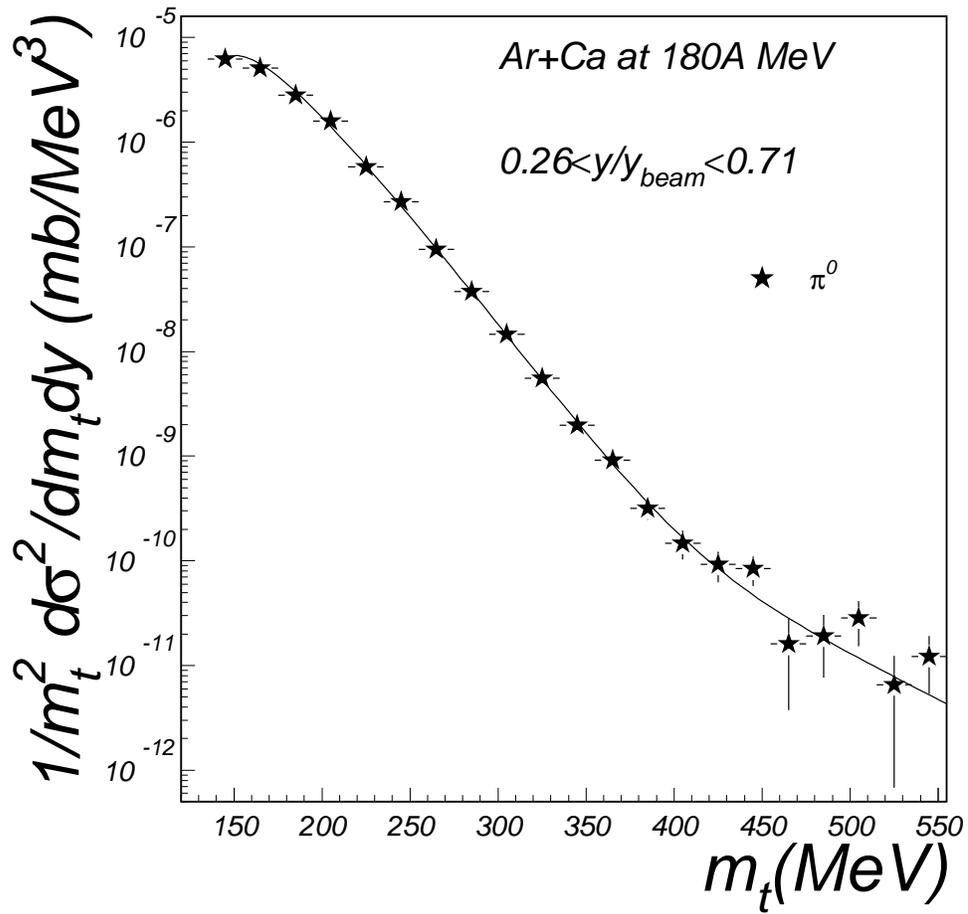}}
  \end{center}
  \label{photon_7}
  \caption{\it Transverse mass distribution of $\pi^\circ$ measured at mid-rapidity for the reaction 
           Ar+Ca at 180A MeV.}
\end{figure}

\begin{figure}
  \begin{center}
   \mbox{\epsfxsize=15cm  \epsfbox{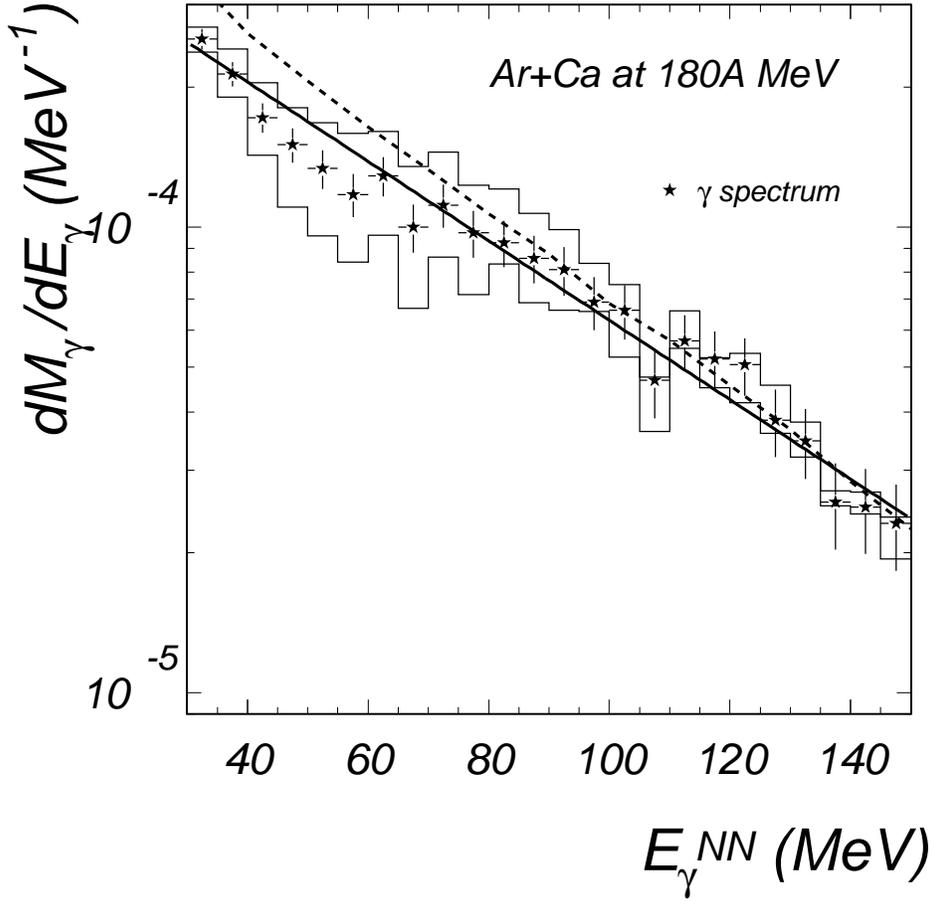}}
  \end{center}
  \label{photon_3}
  \caption{\it Hard-photon {\it spectrum} 
for the reaction Ar+Ca at 180$A$ MeV 
   obtained after subtraction of the $\pi^\circ$ contribution  from the
   inclusive photon spectrum (Fig.1). Histograms represent the systematic 
  error induced by the uncertainty on the anisotropy value of the $\pi^\circ$ decay contribution.
  Solid line represents an exponential fit the measured photon spectrum in the energy range from 30
  to 150 MeV. 
  The dashed line represents a semiclassical transport calculation described in reference \protect\cite{Cass99}
  for the studied system.}
\end{figure}

\begin{figure}
  \begin{center}
   \mbox{\epsfxsize=15cm  \epsfbox{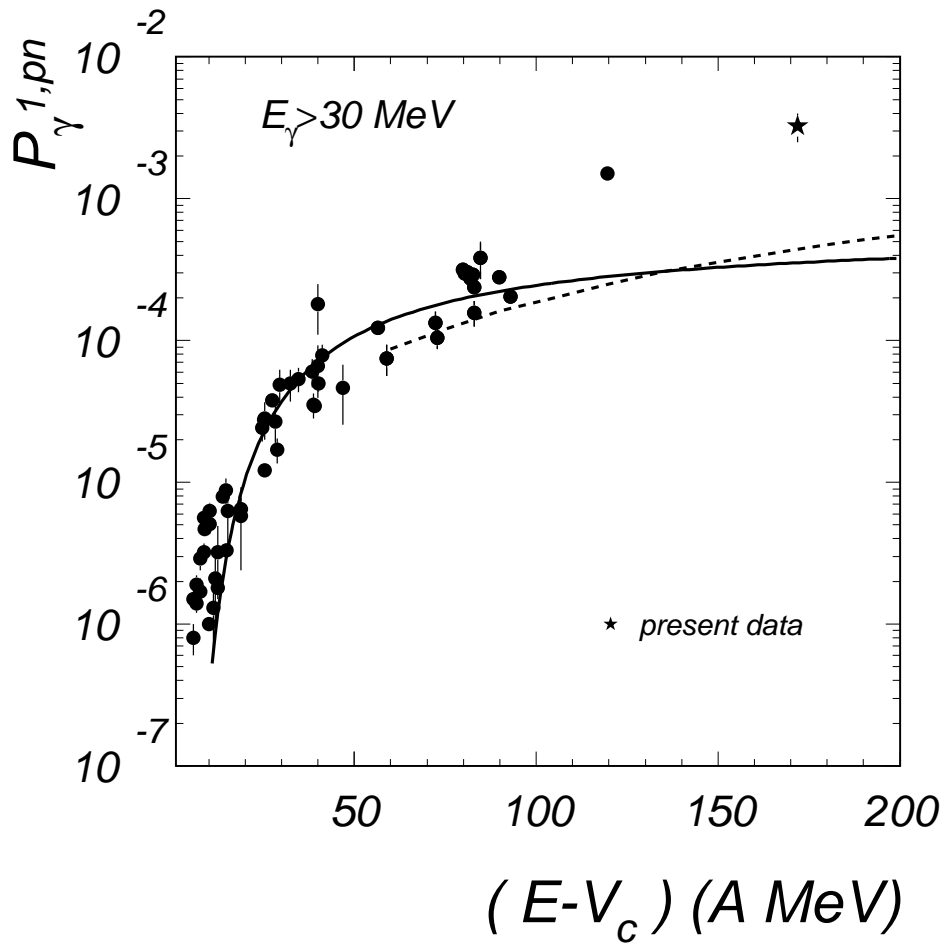}}
  \end{center}
  \label{photon_6}
  \caption{\it Systematics of hard photon production $E_\gamma>30$ MeV (solid line) \protect\cite{Cass90,Clay90,Polt95} 
  compared to the estimation of the photon probability (dashed line) within the simple model described in the text.   
  }
\end{figure}

\end{document}